\documentclass[twocolumn,prb,showpacs]{revtex4}
\usepackage{graphicx}
\usepackage{rotating}
\begin{document}
\title{ New interpretation for energy gap $\Delta$ of the cut-off approximation in the BCS theory of
superconductivity }
\author{Zhidong Hao}
\affiliation{ Department of Physics, University of Science and Technology of China, Hefei, Anhui 230026, China }
\date{ \today }
\begin{abstract}
This paper concerns the solution of the self-consistency equation for energy gap parameter $\Delta_{\bf k}$ in the
BCS theory of superconductivity.  We show that there exists a well-defined relation between the solution for energy
gap parameter amplitude $|\Delta_{\bf k}|$ for a general interaction $V_{{\bf k},{\bf k}'}$ and energy gap $\Delta$
obtained by using the cut-off approximation.  The relation between $|\Delta_{\bf k}|$ and $\Delta$ indicates that
$\Delta$ is a weighted average over $|\Delta_{\bf k}|$ of electronic states within cut-off energy $\xi_c$ around the
Fermi surface. In this interpretation for $\Delta$, $\xi_c$ is not a property of $V_{{\bf k},{\bf k}'}$, but a
parameter specifying the energy range within which the weighted average over $|\Delta_{\bf k}|$ is taken.  We show
that the proper choice for the value of $\xi_c$ is only a few $k_BT_c$ (i.e., $\xi_c/k_BT_c$ is about 3 or 4). We
also show that the cut-off approximation, even with $\xi_c/k_BT_c=\infty$, is a good approximation when it is used to
calculate quantities such as the condensation energy and the specific heat, but it leads to significant
overestimation for the Josephson critical current density of a Josephson junction if $\xi_c/k_BT_c \gg 1$ is assumed.
\end{abstract}
 \maketitle

In the BCS theory of superconductivity,\cite{bcs,bogo} the superconducting state is characterized by the existence of
energy gap parameter $\Delta_{\bf k}$ in quasi-particle excitation energy $E_{\bf k} = \sqrt{\xi_{\bf
k}^2+|\Delta_{\bf k}|^2}$ (where $\xi_{\bf k}$ is the normal state electronic energy, measured relative to the Fermi
level). Energy gap parameter $\Delta_{\bf k}$ is determined self-consistently via the equation
\begin{equation}
\Delta_{\bf k} = -\sum_{\bf k'} V_{{\bf k},{\bf k'}} \frac{\tanh(E_{\bf k'}/2k_BT)}{2E_{\bf k'}}\Delta_{\bf k'},
\label{gap-eq}
\end{equation}
where $V_{{\bf k},{\bf k'}} $ is the pairing interaction matrix element.

In principle, once $\Delta_{\bf k}$ is determined, thermodynamic quantities in the superconducting state can be
quantitatively calculated as functions of temperature $T$ by starting with the diagonalized
Hamiltonian\cite{bcs,bogo,mu}
\begin{equation}
\hat{H} = \sum_{\bf k} \left[ U_{\bf k} + E_{\bf k} \left( \gamma^\dagger_{{\bf k}\uparrow}\gamma_{{\bf k}\uparrow} +
\gamma^\dagger_{-{\bf k}\downarrow}\gamma_{-{\bf k}\downarrow} \right) \right],    \label{H}
\end{equation}
where $U_{\bf k} = \xi_{\bf k} - E_{\bf k} + |\Delta_{\bf k}|^2\tanh(E_{\bf k}/2k_BT)/2E_{\bf k}$, and
$\gamma^\dagger_{{\bf k}\sigma}$ and $\gamma_{{\bf k}\sigma}$ are the Fermi operators for quasi-particles in the
superconducting state.\cite{bcs,bogo}

In general, $\Delta_{\bf k}$ is a complex quantity, i.e., $\Delta_{\bf k} = |\Delta_{\bf k}|e^{i\theta_{\bf k}}$, and
both amplitude $|\Delta_{\bf k}|$ and phase $\theta_{\bf k}$ can be wave vector {\bf k} dependent. The cut-off
approximation,\cite{bcs} in which $V_{{\bf k},{\bf k}'}$ is approximated by
\begin{equation}
V_{{\bf k},{\bf k}'} = \left\{
\begin{array}{ll}
 -V < 0 & \text{ if } |\xi_{\bf k}|, |\xi_{{\bf k}'}| < \xi_c \\
 0  & \text{ otherwise}
\end{array} \right.               \label{cut-off-V}
\end{equation}
and $\theta_{\bf k} = \theta$ is assumed to be a constant (which can be arbitrary), suppresses the {\bf k}-dependence
of $\Delta_{\bf k}$ so that
\begin{equation}
|\Delta_{\bf k}| = \left\{
\begin{array}{ll}
\Delta \ge 0 & \text{if } |\xi_{\bf k}| < \xi_c \\
0 & \text{otherwise}
\end{array}
\right.         \label{cut-off-gap}
\end{equation}
and Eq. (\ref{gap-eq}) becomes
\begin{equation}
1 = V \sum_{|\xi_{\bf k}| < \xi_c} \frac{\tanh\left(\sqrt{\xi_{\bf k}^2 + \Delta^2}/2k_BT \right)}{2 \sqrt{\xi_{\bf
k}^2 + \Delta^2}}.  \label{cut-off-gap-eq}
\end{equation}

Cut-off energy $\xi_c$ was thought to be of the same order as Debye energy $\hbar\omega_D$, i.e., $\xi_c/k_BT_c
\simeq \hbar\omega_D/k_BT_c\gg 1$,\cite{bcs} and $\xi_c/k_BT_c=\infty$ was often assumed in practical calculation of
various quantities.\cite{bcs,mu}

Major quantitative results of the BCS theory were first derived by using the cut-off approximation.\cite{bcs,mu}
Despite the fact that the approximation is oversimplified, the quantitative results have shown, in general, good
agreement with experiments on a variety of (conventional) superconductors.\cite{bs61,parks69}  However, there are
also noteworthy discrepancies. An example is that the predicted value\cite{ab63} for the magnitude of the Josephson
critical current density of a Josephson junction is much too large compared to what experimentally
observed,\cite{ar63,fiske64,yanson64,josephson69} even though the prediction for the temperature dependence of the
normalized Josephson critical current density has been found to be in excellent agreement with experiments.

We have derived in Ref. \onlinecite{hao93} a solution for energy gap parameter amplitude $|\Delta_{\bf k}|$ for a
general interaction $V_{{\bf k},{\bf k}'}$.  The solution for $|\Delta_{\bf k}|$ shows that reduced energy gap
parameter amplitude $|\Delta_{\bf k}|/k_BT_c$ is a function only of reduced variables $|\xi_{\bf k}|/k_BT_c$ and
$T/T_c$, which contains no explicit $V_{{\bf k},{\bf k}'}$-dependence.  The solution also shows that $|\Delta_{\bf
k}|$ is appreciable only for energies within a few $k_BT_c$ around the Fermi level. This latter feature of
$|\Delta_{\bf k}|$ is very different from what one would expect from the cut-off approximation if $\xi_c/k_BT_c\gg 1$
is assumed. Despite this difference, as we have shown in Ref. \onlinecite{hao96}, the results for thermodynamic
critical magnetic field $H_c(T)$, specific heat $C(T)$ and normalized Josephson critical current density
$I_c(T)/I_c(0)$, obtained by using the solution of Ref. \onlinecite{hao93} for $|\Delta_{\bf k}|$, are not much
different from those obtained by using the cut-off approximation (with $\xi_c/k_BT_c=\infty$).  However, there is one
significant difference: the value of $I_c(0)$, obtained by using the solution of Ref. \onlinecite{hao93} for
$|\Delta_{\bf k}|$, is only about a third of that obtained by using the cut-off approximation (with
$\xi_c/k_BT_c=\infty$). The reason behind these is further analyzed and made clear in this paper.

In the following, we show that there exists a well-defined relation between the solution for $|\Delta_{\bf k}|$
obtained in Ref. \onlinecite{hao93} for a general interaction $V_{{\bf k},{\bf k}'}$ and energy gap $\Delta$ of the
cut-off approximation.  The relation between $|\Delta_{\bf k}|$ and $\Delta$ indicates that $\Delta$ is a weighted
average over $|\Delta_{\bf k}|$ of electronic states within $\xi_c$ around the Fermi surface. In this interpretation
for $\Delta$, cut-off energy $\xi_c$ is not a property of the interaction, but a parameter specifying the energy
range within which the weighted average over $|\Delta_{\bf k}|$ is taken.  We show that the proper choice for the
value of $\xi_c$ is only a few $k_BT_c$ (i.e., $\xi_c/k_BT_c$ is about 3 or 4). We also show that the cut-off
approximation, even with $\xi_c/k_BT_c=\infty$, is a good approximation when it is used to calculate quantities such
as condensation energy $H_c^2(T)/8\pi$, specific heat $C(T)$ and normalized Josephson critical current density
$I_c(T)/I_c(0)$, but it leads to significant overestimation for the magnitude of the Josephson critical current
density if $\xi_c/k_BT_c \gg 1$ is assumed.

As we have shown in Refs. \onlinecite{hao93} and \onlinecite{hao96}, the following first order differential equation
for $|\Delta_{\bf k}(T)|$ holds:
\begin{equation}
\sum_{\bf k} |\Delta_{\bf k}|^2 \frac{d}{dT}\left[\frac{\tanh\left(\sqrt{\xi_{\bf k}^2+|\Delta_{\bf k}|^2}/2k_BT
\right)} {\sqrt{\xi_{\bf k}^2+|\Delta_{\bf k}|^2}}\right] = 0.     \label{dgap}
\end{equation}
This equation can be derived from Eq. (\ref{gap-eq}) by first operating $d/dT$ on Eq. (\ref{gap-eq}), and then
multiplying the resulting equation by $(\Delta^{\star}_{\bf k}/2E_{\bf k})\tanh(E_{\bf k}/2k_BT)$ and taking
summation over {\bf k}.  It can also be derived by calculating entropy $S$ from diagonalized Hamiltonian $\hat{H}$ of
Eq. (\ref{H}), and letting the resulting expression for $S$ to be the same as the standard expression expected for a
system of Fermions.\cite{hao93,hao96}

Note that interaction $V_{{\bf k},{\bf k}'}$ and phase $\theta_{\bf k}$ do not appear explicitly in Eq. (\ref{dgap}).
Instead, critical temperature $T_c$ is involved through the condition that $|\Delta_{\bf k}| = 0$ at $T=T_c$.  This
indicates that $|\Delta_{\bf k}|$ depends on $V_{{\bf k},{\bf k}'}$ only implicitly via $T_c$.

To see how critical temperature $T_c$ depends on interaction $V_{{\bf k},{\bf k}'}$, we turn to Eq. (\ref{gap-eq}).
In the limit of $T \rightarrow T_c$, we have $|\Delta_{\bf k}| \rightarrow 0$ so that Eq. (\ref{gap-eq}) can be
linearized, and we have an eigenvalue problem:
\begin{equation}
\Delta_{\bf k} = - \sum_{{\bf k}'} V_{{\bf k},{\bf k}'} \frac{\tanh(|\xi_{{\bf k}'}|/2k_BT_c)}{2|\xi_{{\bf k}'}|}
\Delta_{{\bf k}'}. \label{tc-eq}
\end{equation}
In principle, critical temperature $T_c$ and phase $\theta_{\bf k}$ are determined by solving this eigenvalue problem
for given interaction $V_{{\bf k},{\bf k}'}$ and electronic energy spectrum $\xi_{\bf k}$.

We turn back to Eq. (\ref{dgap}) to consider how a solution for $|\Delta_{\bf k}|$ can be obtained. Clearly,
\begin{equation}
\frac{\tanh\left(\sqrt{\xi_{\bf k}^2 + |\Delta_{\bf k}|^2}/2k_BT\right)} {\sqrt{\xi_{\bf k}^2 + |\Delta_{\bf k}|^2}}
= \frac{\tanh(|\xi_{\bf k}|/2k_BT_c)}{|\xi_{\bf k}|}         \label{my-solution}
\end{equation}
is a solution of Eq. (\ref{dgap}).  This equation, which was previously obtained in Ref. \onlinecite{hao93}, is an
implicit solution for $|\Delta_{\bf k}|$ as a function of $|\xi_{\bf k}|$ and $T$ for given $T_c$, and satisfies the
condition that $|\Delta_{\bf k}| = 0$ at $T=T_c$.

However, Eq. (\ref{my-solution}) is not the only possible solution of Eq. (\ref{dgap}).  Actually, as one can see,
Eq. (\ref{dgap}) can have infinite number of solutions.  For example, the solution of the form of Eq.
(\ref{cut-off-gap}) in the case of the cut-off approximation is also a solution of Eq. (\ref{dgap}). This can be seen
by substituting Eq. (\ref{cut-off-gap}) into Eq. (\ref{dgap}) to obtain
\begin{equation}
\sum_{|\xi_{\bf k}|<\xi_c} \frac{d}{dT}\left[ \frac{\tanh\left(\sqrt{\xi_{\bf k}^2+\Delta^2}/2k_BT\right)}
{\sqrt{\xi_{\bf k}^2+\Delta^2}}\right] = 0, \label{dgap2}
\end{equation}
and noticing that this equation can also be obtained from Eq. (\ref{cut-off-gap-eq}) by operating $d/dT$ on it.
Similarly, a solution of the form $\Delta_{\bf k}=\Delta\omega_{\bf k}$, as in the case of a separable interaction
$V_{\bf k,k'} = -V \omega_{\bf k} \omega_{\bf k'}$, is also a solution of Eq. (\ref{dgap}).  We therefore need an
additional constraint so that $|\Delta_{\bf k}(T)|$ can be uniquely determined.

Note that diagonalized Hamiltonian $\hat{H}$ of Eq. (\ref{H}) is $T$-dependent because of its dependence on
$|\Delta_{\bf k}(T)|$. This implies the existence of an additional self-consistency constraint, which, as we will see
in the following, allows unique determination of $|\Delta_{\bf k}(T)|$.

As we discussed in Ref. \onlinecite{hao93}, since diagonalized Hamiltonian $\hat{H}$ describes a set of independent
quasi-particle excitations, there should be no coupling (except pair correlation) between the quasi-particle
excitations. Therefore, we expect the thermal energy and the entropy associated with each pair of $({\bf
k}\!\uparrow,{\bf -k}\!\downarrow)$ excitations to be
\begin{equation}
\varepsilon_{\bf k} = U_{\bf k}+2f_{\bf k}E_{\bf k} \label{ek}
\end{equation}
and
\begin{equation}
S_{\bf k} = -2k_B\left[f_{\bf k}\ln f_{\bf k} + (1-f_{\bf k})\ln (1-f_{\bf k})\right], \label{Sk}
\end{equation}
respectively [where $f_{\bf k} = (e^{E_{\bf k}/k_BT}+1)^{-1}$ is the Fermi function].  However, as compared to these
standard expressions for $\varepsilon_{\bf k}$ and $S_{\bf k}$, those derived from diagonalized Hamiltonian $\hat{H}$
of Eq. (\ref{H}) contain additional terms involving $dU_{\bf k}/dT$, $dE_{\bf k}/dT$ and $df_{\bf k}/dT$.  Letting
the sum of the additional terms in each expression to be zero, one gets a first order differential equation for
$|\Delta_{\bf k}(T)|$, of which the solution satisfying the condition $|\Delta_{\bf k}(T_c)|=0$ is Eq.
(\ref{my-solution}).\cite{hao93}  Namely, the solution for $|\Delta_{\bf k}(T)|$ given by Eq. (\ref{my-solution}) is
the only solution that both satisfies the original self-consistency equation [Eq. (\ref{gap-eq})] and ensures that
Eqs. (\ref{ek}) and (\ref{Sk}) hold.  [A complete solution for $\Delta_{\bf k}$ is therefore a combination of the
solutions of Eq. (\ref{tc-eq}) for $T_c$ and $\theta_{\bf k}$ and the solution of Eq. (\ref{my-solution}) for
$|\Delta_{\bf k}|$.]

There is a well defined relation between the quantity $\Delta$ of the cut-off approximation and $|\Delta_{\bf k}|$
[hereafter, $|\Delta_{\bf k}|$ means the solution for the energy gap parameter amplitude given by Eq.
(\ref{my-solution})].  This can be seen as follows.

Note that Eq. (\ref{dgap2}) is a first order differential equation for $\Delta(T)$, of which the solution satisfying
the condition $\Delta(T_c)=0$ is
\begin{widetext}
\begin{equation}
\sum_{|\xi_{\bf k}| < \xi_c} \left[ \frac{\tanh\left(\sqrt{\xi_{\bf k}^2 + \Delta^2} / 2k_BT \right)} {\sqrt{\xi_{\bf
k}^2 + \Delta^2}} - \frac{\tanh(|\xi_{\bf k}|/2k_BT_c)} {|\xi_{\bf k}|} \right] = 0.  \label{cut-off-solution}
\end{equation}
By substituting Eq. (\ref{my-solution}) into the above equation, we obtain
\begin{equation}
\sum_{|\xi_{\bf k}| < \xi_c} \left[ \frac{\tanh\left(\sqrt{\xi_{\bf k}^2 + \Delta^2} / 2k_BT\right)} {\sqrt{\xi_{\bf
k}^2+\Delta^2}} - \frac{\tanh\left(\sqrt{\xi_{\bf k}^2+|\Delta_{\bf k}|^2} / 2k_BT\right)} {\sqrt{\xi_{\bf k}^2 +
|\Delta_{\bf k}|^2}} \right] = 0.   \label{definition}
\end{equation}
This equation defines $\Delta$ as a weighted average over $|\Delta_{\bf k}|$ of electronic states with $|\xi_{\bf k}|
< \xi_c$; we therefore can so interpret $\Delta$. In this interpretation for $\Delta$, $\xi_c$ is not a property of
the interaction, but a parameter specifying the energy range within which the weighted average over $|\Delta_{\bf
k}|$ is taken.

We next examine how $\Delta$ depends on $\xi_c$.  From Eq. (\ref{dgap2}), by making the usual substitution
$\sum_{|\xi_{\bf k}| < \xi_c} \rightarrow N(0)\int_0^{\xi_c}d\xi$ [where $N(0)$ is the density of states at the Fermi
level] and a rearrangement, we obtain the following expression:
\begin{equation}
\frac{d\left(\Delta^2\right)}{dT} = - \frac{ \int_0^{\xi_c} d\xi \left[ \mbox{sech}^2 \left( \sqrt{\xi^2+\Delta^2}/2T
\right) /T^2 \right]} { \int_0^{\xi_c} d\xi \left[ \tanh\left(\sqrt{\xi^2+\Delta^2}/2T\right) / \left( \xi^2 +
\Delta^2 \right)^{3/2} - \mbox{sech}^2\left(\sqrt{\xi^2+\Delta^2}/2T\right) / 2T\left(\xi^2+\Delta^2\right)\right] },
                \label{diff-eq-2}
\end{equation}
\end{widetext}
where energies are measured in units of $k_BT_c$. With this expression for $d(\Delta^2)/dT$ and the initial value
$\Delta^2=0$ at $T=1$ (temperature $T$ is measured in unit of $T_c$), we can numerically calculate $\Delta(T)$ for
arbitrary $\xi_c$ by using the Runge-Kutta method.\cite{conte80} The integrals involved in the expression for
$d(\Delta^2)/dT$ are calculated by using the Simpson method.\cite{conte80}

We show in Fig. 1 the energy dependence of the gap parameter amplitude at $T=0$. The results for $\Delta(0)$ obtained
for $\xi_c/k_BT_c = 1$, 2, 3.37, 5, 10 and $\infty$ are shown as plots (a), (b), (c), (d), (e) and (f), respectively.
(Here the number 3.37 is special, as we will see later.) Plot (g) shows $|\Delta_{\bf k}(0)|$ versus $|\xi_{\bf k}|$.

Also note that if a different cut-off range is assumed so that $|\Delta_{\bf k}|=\Delta$ for $\xi_{c1} < |\xi_{\bf
k}| < \xi_{c2}$ and zero otherwise, then $\Delta$ is simply a weighted average over $|\Delta_{\bf k}|$ of electronic
states with $\xi_{\bf k}$ in the range $\xi_{c1}<|\xi_{\bf k}|<\xi_{c2}$. As examples, the results for
$(\xi_{c1}/k_BT_c, \xi_{c2}/k_BT_c) = (6, 7)$, (7, 8) and (8, 9) are shown as plots (h), (i) and (j), respectively,
in Fig. 1.

Comparing plots (a)-(f) and (h)-(j) with plot (g), it is evident that $\Delta$ is indeed a weighted average over
$|\Delta_{\bf k}|$ of energies within a specific range.  This relation between $\Delta$ and $|\Delta_{\bf k}|$ is
mathematically expressed by Eq. (\ref{definition}).  We can also see from Fig. 1 that electronic states with lower
$|\xi_{\bf k}|$ contribute with larger weights to the average. The fact that the difference between the values of
$\Delta(0)$ for $\xi_c/k_BT_c = 10$ [plot (e)] and $\xi_c/k_BT_c = \infty$ [plot (f)] is less than $1\%$ indicates
that contributions from electronic states with $|\xi_{\bf k}|/k_BT_c \gg 1$ are negligibly small.

\begin{figure}
\includegraphics{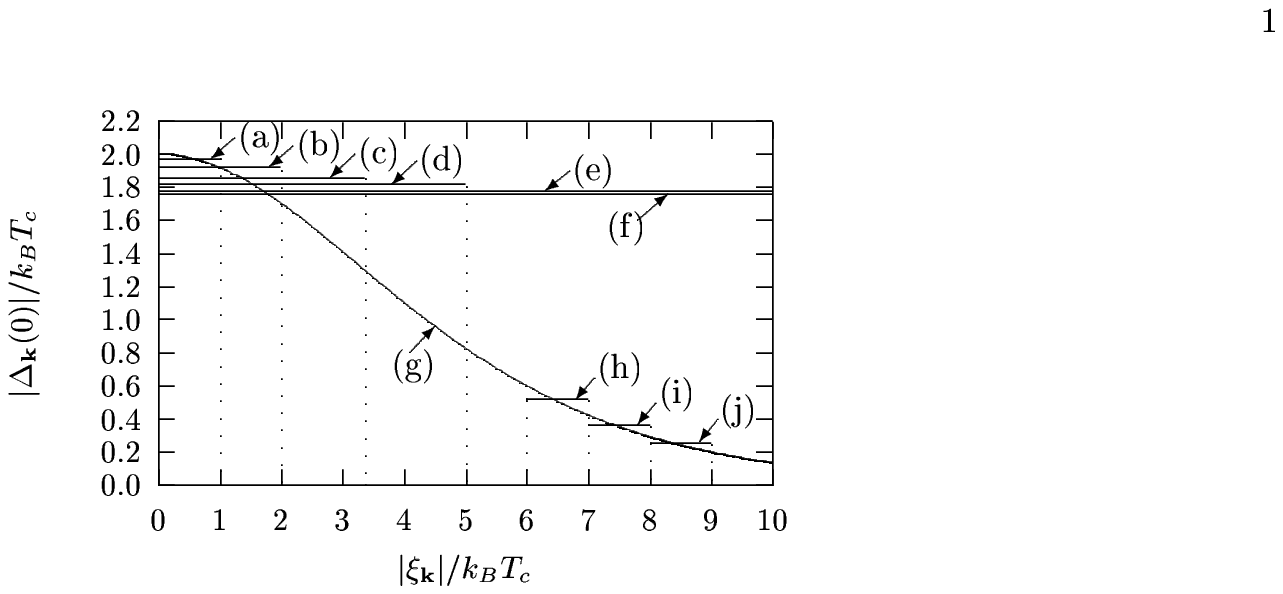}
\caption{Energy dependence of the energy gap parameter amplitude at $T\!=\!0$. Plots (a)-(f): $\Delta(0)/k_BT_c =
1.975$, 1.922, 1.859, 1.816, 1.778 and 1.764 for $\xi_c/k_BT_c = 1$, 2, 3.37, 5, 10 and $\infty$, respectively.  Plot
(g): $|\Delta_{\bf k}(0)|$ versus $|\xi_{\bf k}|$.  Plots (h)-(j): $\Delta(0)/k_BT_c = 0.517$, 0.364 and 0.255 for
$(\xi_{c1}/k_BT_c, \xi_{c2}/k_BT_c) = (6, 7)$,  (7, 8) and (8, 9), respectively.} \label{figure1}
\end{figure}

The minimum single quasi-particle excitation energy is $E_{{\bf k},\text{min}}=|\Delta_{{\bf k}_F}|$ (where ${\bf
k}_F$ is a Fermi wave vector).  We have $|\Delta_{{\bf k}_F}(0)|/k_BT_c = 2$. The result $\Delta(0)/k_BT_c=1.764$ for
$\xi_c/k_BT_c=\infty$ [plot (f) in Fig. 1] is about $88\%$ of $|\Delta_{{\bf k}_F}(0)|/k_BT_c$\,.

In Fig. 2, we compare the $T$-dependence of normalized energy gap $|\Delta_{{\bf k}_F}(T)|/|\Delta_{{\bf k}_F}(0)|$
with that of $\Delta(T)/\Delta(0)$. Four practically indistinguishable curves are shown in Fig. 2: $|\Delta_{{\bf
k}_F}(T)|/|\Delta_{{\bf k}_F}(0)|$; and $\Delta(T)/\Delta(0)$ for $\xi_c/k_BT_c = 1$, 100 and $\infty$ of the cut-off
approximation [$\Delta(T)/\Delta(0)$ for $\xi_c/k_BT_c = \infty$ was previously calculated by
M\"{u}hlschlegel\cite{mu}]. As shown in Fig. 2, $\Delta(T)/\Delta(0)$ is practically $\xi_c$-independent, and
practically the same as $|\Delta_{{\bf k}_F}(T)|/|\Delta_{{\bf k}_F}(0)|$.

\begin{figure}
\includegraphics{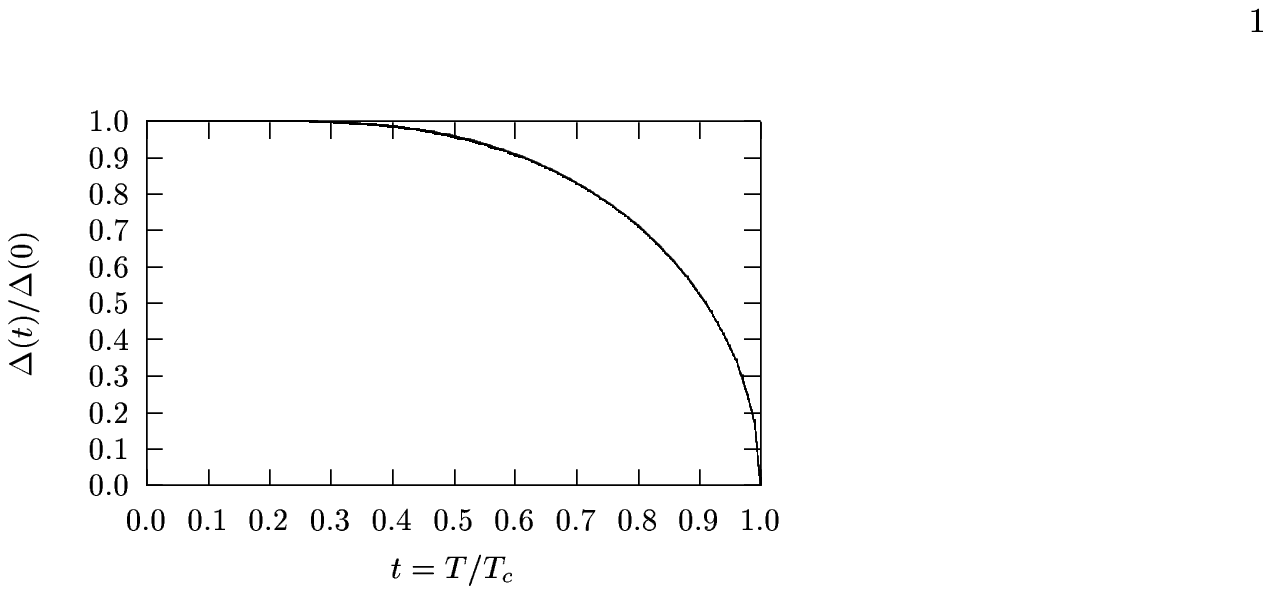}
\caption{Temperature dependence of the normalized energy gap. Four practically indistinguishable curves are plotted
in this figure: $|\Delta_{{\bf k}_F}(t)|/|\Delta_{{\bf k}_F}(0)|$; and $\Delta(t)/\Delta(0)$ for $\xi_c/k_BT_c = 1$,
100 and $\infty$. } \label{figure2}
\end{figure}

\begin{figure}
\includegraphics{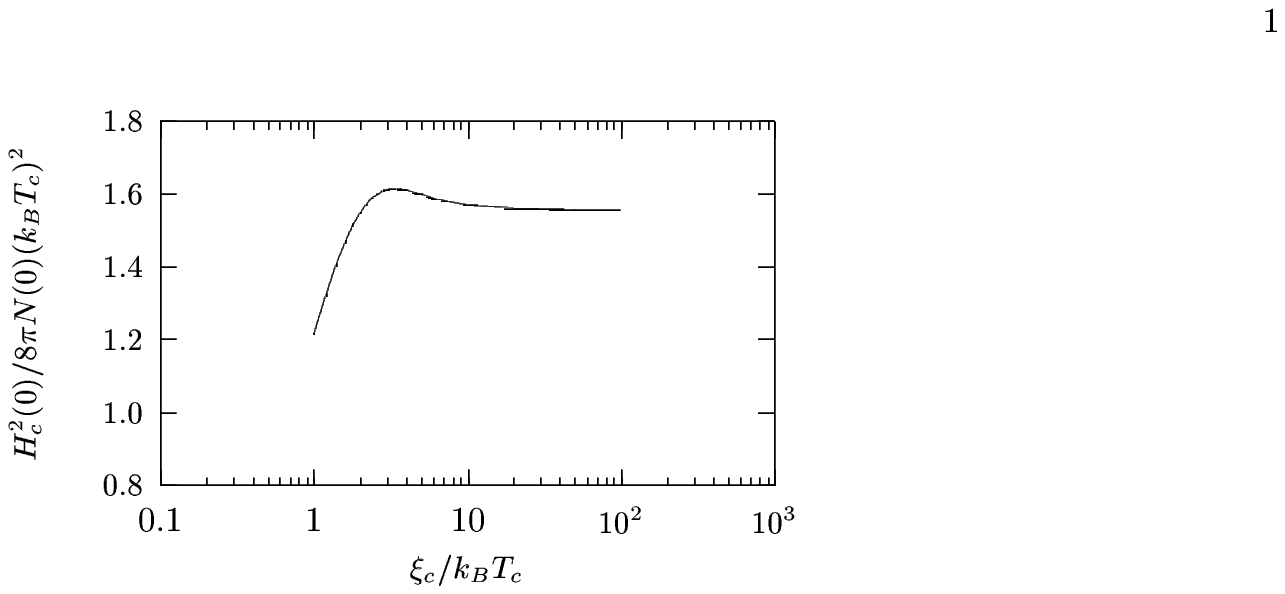}
\caption{ Zero-temperature condensation energy $H_c^2(0)/8\pi$ versus cut-off energy $\xi_c$. The maximum
$H_c^2(0)/8\pi=1.613$ [in unites of $N(0)(k_BT_c)^2$] is located at $\xi_c/k_BT_c=3.37$; $H_c^2(0)/8\pi = 1.556$ for
$\xi_c/k_BT_c=\infty$. For comparison, the results calculated by using $|\Delta_{\bf k}|$ is $H_c^2(0)/8\pi = 1.645$.
} \label{figure3}
\end{figure}

We next examine how other quantities such as thermodynamic critical magnetic field $H_c(T)$, specific heat $C(T)$ and
Josephson critical current density $I_c(T)$ depend on cut-off energy $\xi_c$ when they are calculated by using the
cut-off approximation.

Once $\Delta(T)$ is obtained, quantities such as $H_c(T)$, $C(T)$ and $I_c(T)$ can be calculated
straightforwardly.\cite{noteHCI} In Fig. 3, we show the $\xi_c$-dependence of zero-temperature condensation energy
$H_c^2(0)/8\pi$.  The $H_c^2(0)/8\pi$-versus-$\xi_c$ curve shows a maximum $H_c^2(0)/8\pi = 1.613$ [in unites of
$N(0)(k_BT_c)^2$] at $\xi_c/k_BT_c = 3.37$.  In the limit of $\xi_c/k_BT_c=\infty$, $H_c^2(0)/8\pi=1.556$.  Note that
these two values of $H_c^2(0)/8\pi$ are only about 2$\%$ and 5$\%$, respectively, smaller than the result
$H_c^2(0)/8\pi=1.645$ calculated by using $|\Delta_{\bf k}|$.

In Fig. 4, we show the $T$-dependence of thermodynamic critical magnetic field $H_c$. The results are plotted as
deviations from the $1-(T/T_c)^2$ law. For comparison, the result calculated by using $|\Delta_{\bf k}|$ and the
results calculated by using the cut-off approximation for several different values of $\xi_c$ are plotted in the
figure.

\begin{figure}
\includegraphics{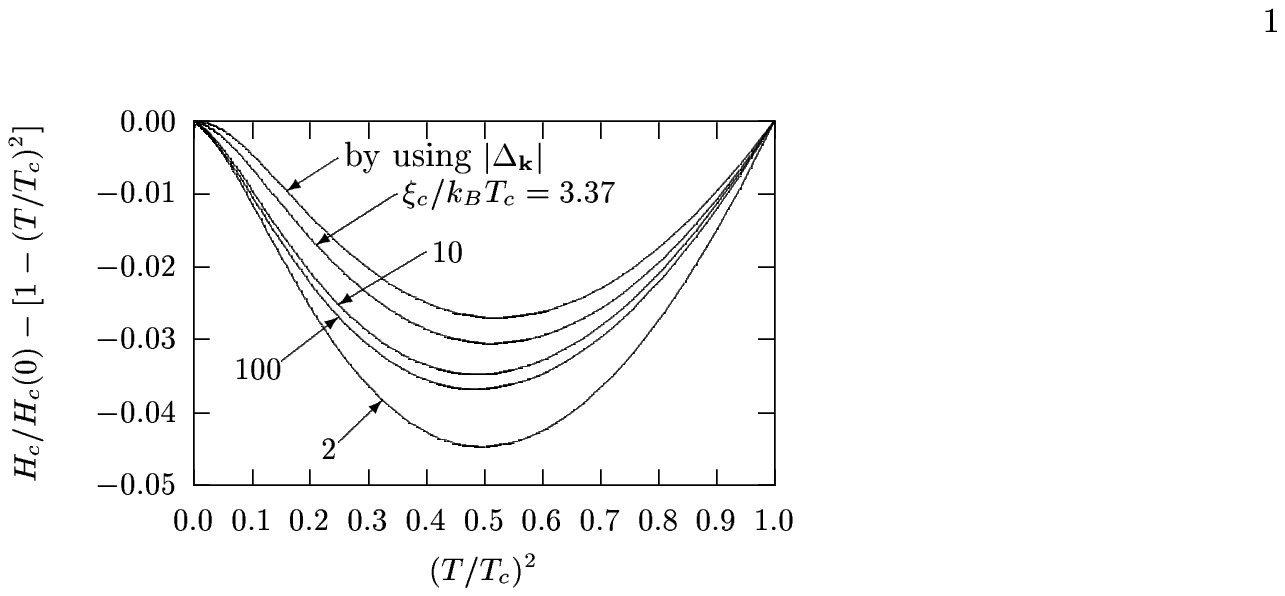}
\caption{Deviation function $D_H = H_c(T)/H_c(0)-[1-(T/T_c)^2]$ versus $(T/T_c)^2$, calculated by using $|\Delta_{\bf
k}|$ and by using the cut-off approximation with different values of $\xi_c/k_BT_c$ as indicated on the curves. }
\label{figure4}
\end{figure}

\begin{figure}
\includegraphics{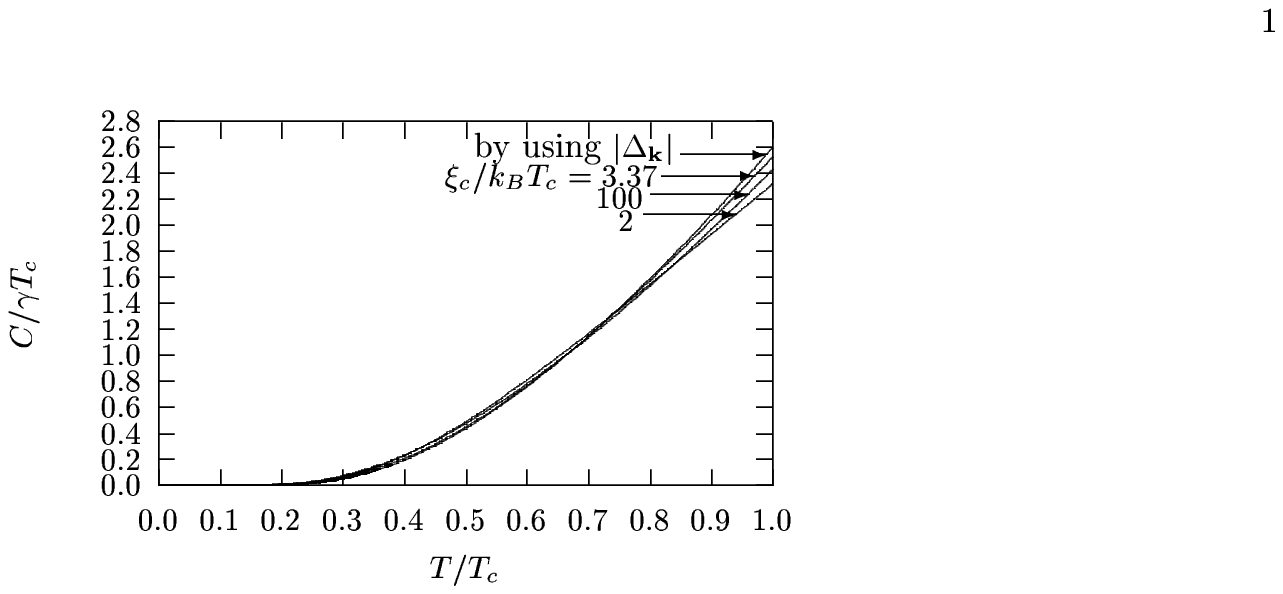}
\caption{Electronic specific heat $C$ versus temperature $T$, calculated by using $|\Delta_{\bf k}|$ and by using the
cut-off approximation with different values of $\xi_c/k_BT_c$ as indicated on the curves. The $C(T_c)/\gamma T_c$
values for the four curves shown in the figure are 2.597, 2.525, 2.482 and 2.318, respectively. } \label{figure5}
\end{figure}

In Fig. 5, we show results for the $T$-dependence of electronic specific heat $C$. For comparison, the result
calculated by using $|\Delta_{\bf k}|$ and the results calculated by using the cut-off approximation for several
different values of $\xi_c$ are plotted in the figure as $C/\gamma T_c$ versus $T/T_c$ [where $\gamma =
(2\pi^2/3)k_B^2N(0)$]. At $T=T_c$, we have $C/\gamma T_c=2.597$ by using $|\Delta_{\bf k}|$, and $C/\gamma
T_c=2.525$, $2.482$ and $2.318$ by using the cut-off approximation for $\xi_c/k_BT_c=3.37$, $100$ and $2$,
respectively.  The result $C(T_c)/\gamma T_c=2.426$ for $\xi_c/k_BT_c=\infty$ was previously obtained by
M\"{u}hlschlegel.\cite{mu}

From Figs. 3-5, we can see that, for calculating $H_c$ and $C$, the cut-off approximation gives results that are not
much different from those obtained by using $|\Delta_{\bf k}|$. The approximation is optimized when
$\xi_c/k_BT_c=3.37$, suggesting the proper choice for $\xi_c$ is about a few $k_BT_c$.\cite{note2} We also note that
the results for $H_c$ and $C$ show only weak $\xi_c$-dependence when $\xi_c$ is a few $k_BT_c$ or larger, so that the
cut-off approximation remains a good approximation even with $\xi_c/k_BT_c\gg 1$. The reason for the weak
$\xi_c$-dependence of $H_c$ and $C$ is that, as one can see from the expressions for $H_c$ and $C$,\cite{noteHCI} the
relevant quantity for $H_c$ and $C$ is quasi-particle excitation energy $E_{\bf k}=\sqrt{\xi_{\bf k}^2 + \Delta^2}$
(or $E_{\bf k}=\sqrt{\xi_{\bf k}^2 + |\Delta_{\bf k}|^2}$\,), which becomes $E_{\bf k}\simeq |\xi_{\bf
k}|=\text{independent of } \Delta$ (or $|\Delta_{\bf k}|$) for $|\xi_{\bf k}|\gg k_BT_c$, so that only electronic
states within a few $k_BT_c$ around the Fermi surface contribute significantly to the difference between the
superconducting and normal states. The situation is different in the case of Josephson critical current density $I_c$
of a Josephson junction, because, as one can see from the expression for $I_c$,\cite{noteHCI} the relevant quantity
for $I_c$ is energy gap parameter $\Delta_{\bf k}$ itself. This leads to a strong $\xi_c$-dependence for $I_c$ when
it is calculated by using the cut-off approximation, as we will see next.

\begin{figure}
\includegraphics{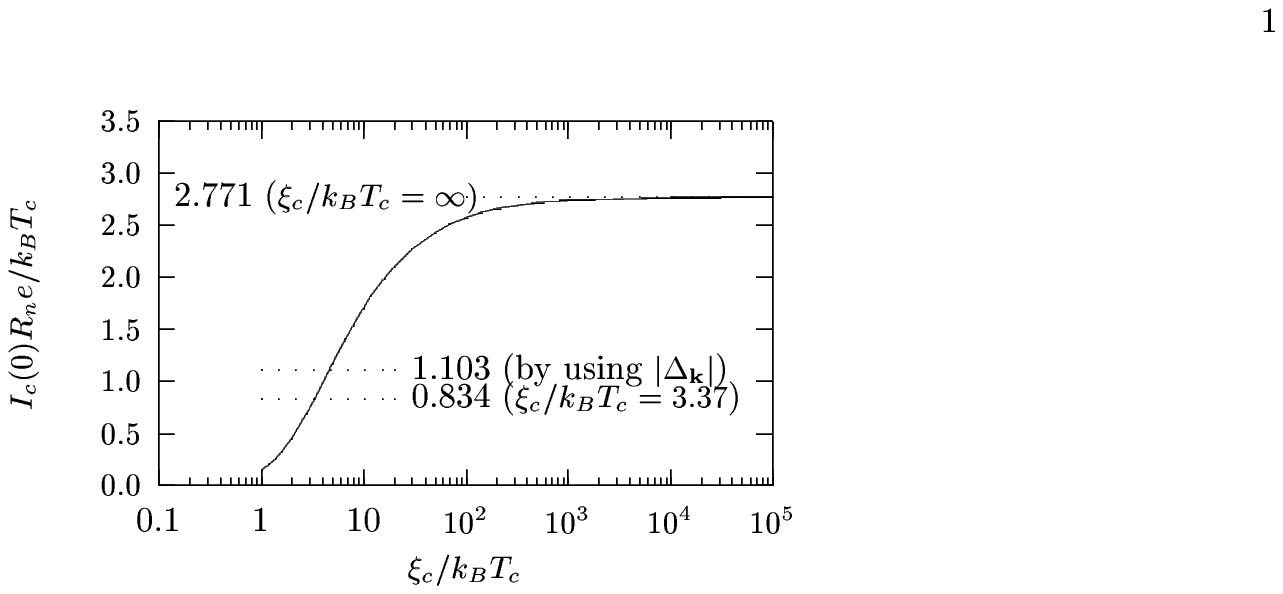}
\caption{The $\xi_c$-dependence of zero-temperature Josephson critical current density $I_c(0)$ of a symmetric
Superconductor-Insulator-Superconductor junction. As indicated in the figure, $I_c(0)=0.834$ and $2.771$ (in units of
$k_BT_c/R_ne$) for $\xi_c/k_BT_c=3.37$ and $\infty$, respectively.  The result obtained by using $|\Delta_{\bf k}|$
is $I_c(0)=1.103$. } \label{figure6}
\end{figure}

\begin{figure}
\includegraphics{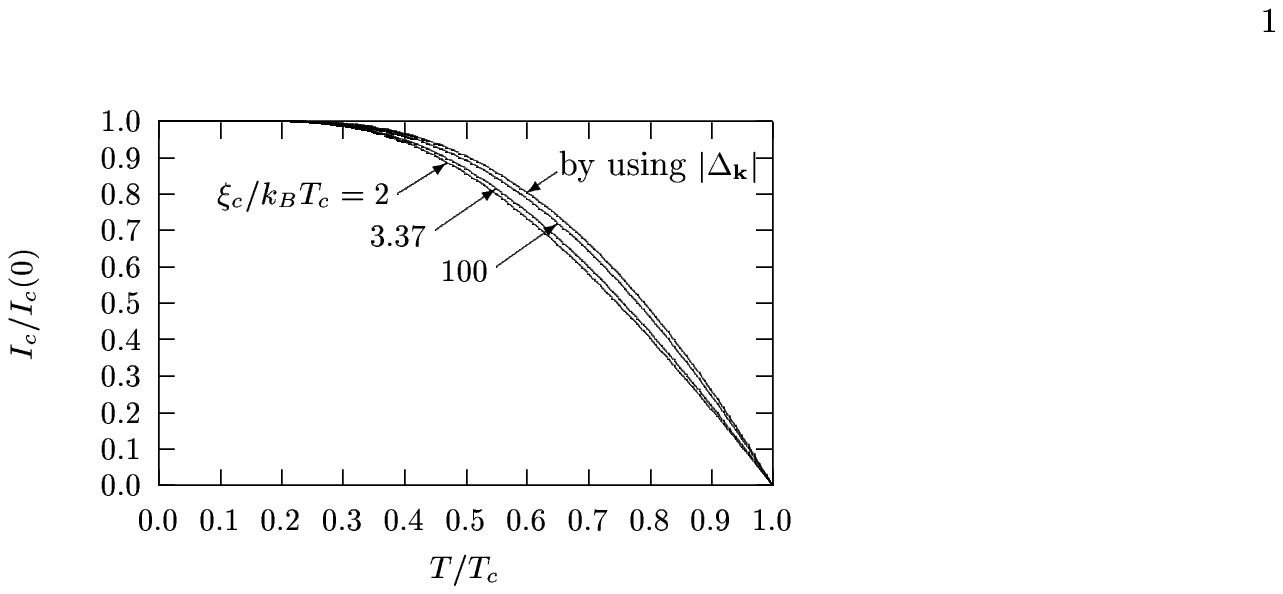}
\caption{ Comparison between the $T$-dependence of normalized Josephson critical current density $I_c(T)/I_c(0)$ of a
symmetric Superconductor-Insulator-Superconductor junction calculated by using $|\Delta_{\bf k}|$ and those by using
the cut-off approximation with different values of $\xi_c/k_BT_c$ as indicated in the figure. } \label{figure7}
\end{figure}

In Fig. 6, we show the $\xi_c$-dependence of zero-temperature Josephson critical current density $I_c(0)$ of a
symmetric Superconductor-Insulator-Superconductor junction. As shown in the figure, $I_c(0)$ is a monotonically
increasing function of $\xi_c$ that does not saturate until $\xi_c/k_BT_c\sim 10^3$. As indicated in the figure, we
have $I_c(0)=0.834$ and $2.771$ (in units of $k_BT_c/R_ne$) for $\xi_c/k_BT_c=3.37$ and $\infty$, respectively (the
result of $I_c$ for $\xi_c/k_BT_c=\infty$ was previously obtained by Ambegaokar and Baratoff\cite{ab63}). For
comparison, the result $I_c(0)=1.103$ calculated by using $|\Delta_{\bf k}|$ is also indicated in the figure, which
is only about 40\% of the result calculated by using the cut-off approximation with $\xi_c/k_BT_c=\infty$. It is
evident from the results shown in Fig. 6 that the cut-off approximation significantly overestimates $I_c$ if
$\xi_c/k_BT_c\gg 1$ is assumed.

In Fig. 7, we show the $T$-dependence of normalized Josephson critical current density $I_c/I_c(0)$ of a symmetric
Superconductor-Insulator-Superconductor junction. For comparison, results calculated by using the cut-off
approximation for several different values of $\xi_c$ and that by using $|\Delta_{\bf k}|$ are plotted in the figure.
As we can see from the figure, for lower values of $\xi_c/k_BT_c$, the difference between the result obtained by
using the cut-off approximation and that by using $|\Delta_{\bf k}|$ is appreciable. However, the difference becomes
much less significant for $\xi_c/k_BT_c\gg 1$.

Experimentally,\cite{ar63,fiske64,yanson64,josephson69,balsamo,sun} the $T$-dependence of $I_c/I_c(0)$ obtained by
using the cut-off approximation with $\xi_c/k_BT_c=\infty$, which is practically indistinguishable from or very close
to that obtained by using $|\Delta_{\bf k}|$, has been well confirmed, but the magnitude of $I_c$ has always been
found to be much too small compared to that predicted by using the cut-off approximation with $\xi_c/k_BT_c=\infty$.
Clearly, as we can see from Figs. 6 and 7, the results obtained by using $|\Delta_{\bf k}|$ can help improve
agreement between theory and experiments.

In summary, we have shown in this paper that energy gap $\Delta$ that was originally obtained by using the cut-off
approximation can be interpreted as a weighted average over $|\Delta_{\bf k}|$ of electronic states within $\xi_c$
around the Fermi surface [here $|\Delta_{\bf k}|$ is the energy gap parameter amplitude for a general interaction
$V_{{\bf k},{\bf k}'}$, and satisfies Eq. (\ref{my-solution})].  In this interpretation for $\Delta$, cut-off energy
$\xi_c$ is not a property of the interaction, but a parameter specifying the energy range within which the weighted
average over $|\Delta_{\bf k}|$ is taken.  We have shown that the proper choice for the value of $\xi_c$ is only a
few $k_BT_c$ (i.e., $\xi_c/k_BT_c$ is about 3 or 4). We have also shown that the cut-off approximation, even with
$\xi_c/k_BT_c=\infty$, is a good approximation when it is used to calculate quantities such as condensation energy
$H_c^2(T)/8\pi$, specific heat $C(T)$ and normalized Josephson critical current density $I_c(T)/I_c(0)$, but it leads
to significant overestimation for the magnitude of the Josephson critical current density if $\xi_c/k_BT_c \gg 1$ is
assumed.


\begin{thebibliography}{}

\bibitem{bcs} J. Bardeen,  L. N. Cooper, and  J. R. Schrieffer, Phys. Rev. {\bf 108}, 1175 (1957).

\bibitem{bogo} N. N. Bogoliubov, Nuovo Cimento {\bf 7}, 794 (1958); Zh. Eksp. Teor. Fiz. {\bf 34}, 58 (1958)
[Sov. Phys. JETP {\bf 7}, 41 (1958)]; J. G. Valatin, Nuovo Cimento {\bf 7}, 843 (1958).

\bibitem{mu} B. M\"{u}hlschlegel, Z. Phys. {\bf 155}, 313 (1959). [English translation in {\it The Theory of
Superconductivity}, edited by N. N. Bogoliubov. Gordon and Breach, Science Publishers, New York (1968).]

\bibitem{bs61} J. Bardeen and  J. R. Schrieffer, in {\it Progress in Low Temperature Physics} (Edited by C. J. Gorter)
Vol. 3, p. 170. North-Holland, Amsterdam (1961).

\bibitem{parks69} R. Meservey and B. B. Schwartz, in {\it Superconductivity} (Edited by R. D. Parks).
Marcel Dekker, New York (1969).

\bibitem{ab63} V. Ambegaokar and A. Baratoff, Phys. Rev. Lett. {\bf 10}, 486 (1963); erratum {\bf 11}, 104 (1963).

\bibitem{ar63} P. W. Anderson and J. M. Rowell, Phys. Rev. Lett. {\bf 10}, 230 (1963).

\bibitem{fiske64} M. D. Fiske, Rev. Mod. Phys. {\bf 36}, 211 (1964).

\bibitem{yanson64} I. K. Yanson, V. M. Svistunov, and I. M. Dmitrenko, Z. Eksp. Teor. Fiz. {\bf 47}, 2091 (1964)
[Sov. Phys. JETP {\bf 20}, 1404 (1965)].

\bibitem{josephson69} B. D. Josephson, in {\it Superconductivity} (Edited by R. D. Parks). Marcel Dekker,
New York (1969).

\bibitem{hao93} Z. Hao, Mod. Phys. Lett. {\bf B7}, 1439 (1993).

\bibitem{hao96} Z. Hao, J. Phys. Chem. Solids {\bf 57}, 1215 (1996).

\bibitem{conte80} S. D. Conte and C. de Boor, {\it Elementary Numerical Analysis: An Algorithmic Approach,}
3rd edition, McGraw-Hill, New York (1980).

\bibitem{noteHCI} See Ref. \onlinecite{hao96} or references cited therein for details about expressions for $H_c$, $C$
and $I_c$. In calculating these quantities, as usual, the substitution $\sum_{\bf k}\rightarrow N(0)\int d\xi$ is
made, and resulting integrals are numerically calculated by using the Simpson method.\cite{conte80}

\bibitem{note2} The value $\xi_c/k_BT_c = 3.37$ is obtained by maximizing condensation energy $H_c^2/8\pi$ at $T=0$.
The optimize value for $\xi_c$ may show some degree of $T$-dependence, but it is not important, as our point here is
to show that the proper choice for $\xi_c$ is only a few $k_BT_c$.

\bibitem{balsamo} E. P. Balsamo, G. Paterno, A. Barone, P. Rissman, and M. Russo, Phys. Rev. B {\b 10}, 1881 (1974).

\bibitem{sun} A. G. Sun, D. A. Gajewski, M. B. Maple, and R. C. Dynes, Phys. Rev. Lett. {\bf 72}, 2267 (1994).


\end{thebibliography}
\end{document}